\documentclass{article}

\usepackage{arxiv}

\usepackage[utf8]{inputenc} 
\usepackage[T1]{fontenc}    
\usepackage{hyperref}       
\usepackage{url}            
\usepackage{amsfonts}       
\usepackage{nicefrac}       
\usepackage{microtype}      

 \usepackage{pgf}
\usepackage{amsmath,amssymb}
\usepackage{graphicx}
\usepackage{environ}
\usepackage{units}
\usepackage{amssymb}
\usepackage{amsmath}
\usepackage{pgf,tikz}
\usepackage{setspace}
\usepackage{ulem}
\usepackage{amsfonts}
\usetikzlibrary{shapes.geometric}
\usetikzlibrary{shapes.arrows}
\usetikzlibrary{arrows,scopes}
\usetikzlibrary{positioning}
\definecolor{lightgray}{rgb}{0.9,0.9,0.9}
\usepackage{upgreek}
 
\title{Lensless hyperspectral imaging by Fourier transform spectroscopy for broadband visible light:  phase retrieval technique}

\author{
  Igor Shevkunov \\
  Faculty of Information Technology and Communication Sciences\\
  Tampere University\\
  Tampere, Korkealunkatu 10 \\
  \texttt{igor.shevkunov@tuni.fi} \\
  \And
Vladimir Katkovnik \\
  Faculty of Information Technology and Communication Sciences\\
  Tampere University\\
  Tampere, Korkealunkatu 10 \\
  \texttt{vladimir.katkovnik@tuni.fi} \\
   \And
 Karen Egiazarian \\
  Faculty of Information Technology and Communication Sciences\\
  Tampere University\\
  Tampere, Korkealunkatu 10 \\
  \texttt{Karen.Egiazarian@tuni.fi} \\
}
\begin{document}
\maketitle

\begin{abstract}
A novel phase retrieval algorithm {\color{black} for broadband} hyperspectral phase imaging from noisy intensity observations is proposed. 
It utilizes advantages of the Fourier Transform spectroscopy in the self-referencing {\color{black} optical setup} and provides, additionally {\color{black} beyond spectral intensity distribution,} reconstruction of the investigated object's phase.
The noise amplification Fellgett's disadvantage is relaxed by  application of sparse wavefront noise filtering embedded in the proposed algorithm.
The algorithm reliability is proved by  {\color{black} simulation tests  and results of physical experiments on transparent objects which demonstrate precise phase imaging and object depth (profile) reconstructions.}
\end{abstract}




\section{Introduction}

Hyperspectral imaging (HSI) is a technique concerned with capturing 3D data where the first two dimensions are {\color{black} spatial 
coordinates defining 2D images (slices of 3D data cube)
    and the third coordinate is a spectral variable. }
{\color{black} This third} dimension adds value to traditional imaging with ability to recover {\color{black} from spectra} extra information about an investigated object, as for example in paintings investigations \cite{HSI_paintings_2016},  disease detection (e.g. Alzheimer's \cite{HSI_alzheimer_natcom_2019},  Parkinson's \cite{HSI_parkinson_2018}), or for food ripeness evaluation \cite{HSI_ripeness2016}.
 
{\color{black} In straightforward approaches for 3D data spectral capturing,} detectors with a number of color filters are used, as for example in \cite{aviris1998}. 
{\color{black}  The hyperspectral digital holography (HSDH) is an alternative approach with a single detector and no color filters,} appeared in the 90's years of the previous century starting from the work \cite{itoh1990}, where  HSDH  has been developed employing principles of incoherent holography and Fourier spectroscopy. 
The algorithm proposed by the authors was implemented in a self-reference optical scheme and performed only amplitude imaging with the phase information lost.

As long as HSI has been used mainly in geoscience and remote sensing \cite{Zhonping_water_HSsensing_1998}, phase information was not necessary for investigators.
 {\color{black} Nowadays,  HSI spreads in the areas of biological and medical applications  \cite{Ba2018,Senlin_HSI_dmd_2017},} where phase becomes crucial and almost vital as it brings additional information about the thickness and refractive index of cells and their behavior (e.g. \cite{Belashov_necrosis_paptosis_2018, PopescuSLIM2018}) without dyeing and additional sample preparations.
To perform phase imaging in HSI, the {\color{black} novel HSDH techniques were developed  with  various  reference beams \cite{Kalenkov2017} }.

 It is possible to retrieve phase in the self-reference setup without any reference beam by modulation one of the object's beams and filtering, which keeps only zero-order diffraction and suppresses all other orders as it is done in  \cite{Kalenkov_selfref_2019}. This solution requires a quite complex system with special lenses and other optical components as well as a projection of imaging on a sensor plane.
 
In this paper, we propose and study a different approach to phase imaging in the HSDH. The system is lensless self-reference and much simpler in implementation as compared with \cite{Kalenkov_selfref_2019}, without lenses and projection on the sensor plane.
To resolve the phase imaging for the object, the original iterative HS phase retrieval algorithm was developed for HS quantitative phase imaging from noisy intensity measurements.
In this way, the complexity of the problem was shifted from the optical hardware components to the software of the algorithm.
The diversity needed for HS phase retrieval is enabled by the joint processing of spectral wavefronts covering a broadband range of wavelengths. 

The developed algorithm and its verification by simulation and physical experiments are the main contributions of this paper. 
The advantage of the proposed solution is the simplicity of the lensless optical setup which provides large field-of-view and makes results free from color aberrations which could be extremely corrupting for HS illumination.
\section{Problem description}
\label{sec:problem}

The self-reference optical scheme assumes that the basic light beam and its phase-shifted copy go through or reflect from the object simultaneously. 
The interference of these two beams is registered at the sensor plane. The phase-shift (phase-delay) is varying and known.
It is usually performed by an interferometric scheme with a moving mirror in one {\color{black} of the two arms, with $N$ steps of size $\Delta z$ covering the  whole phase-delay distance $Z=N\cdot\Delta z$.}

{\color{black} Let $V(\lambda)$ be the complex-valued spectrum of a broadband light beam after free propagation through the object to the sensor plane,  $\lambda\in\Lambda$, where $\Lambda$ is a spectral range of illumination.}
Then the recorded intensity for the mirror-shift distance $z$ in the delay line is calculated as integral over $\Lambda$:
\begin{equation}
J(z)=\int_{\Lambda}\left|V(\lambda)\left(1+\exp (j \dfrac{2 \pi}{ \lambda} z)\right)\right|^{2} d \lambda.
\label{eq:interferog}
\end{equation}
{\color{black} The Fourier transform calculated for $J(z)$ over $z$ 
allows to get the intensity spectrum for $V(\lambda)$:} 
\begin{equation}
    |V(\lambda)|^2=|\mathcal{FT}(J(z))|, \lambda\in\Lambda,
    \label{eq:amplitude}
\end{equation}
where $\mathcal{FT}$ stays for the  Fourier Transform. 
 
 As a result, we obtain a set of squared spectra $|V(\lambda)|^2$ for a set of wavelengths $\Lambda$ at the sensor plane but the phase of $V(\lambda)$ information is lost.

The intensities $J(z)$ calculated over the spectral interval $\Lambda$ are registered by a multi-pixel detector and the same spectral calculations are valid for each pixel of the detector. 
\section{Algorithm}
\label{sec:algorithm}

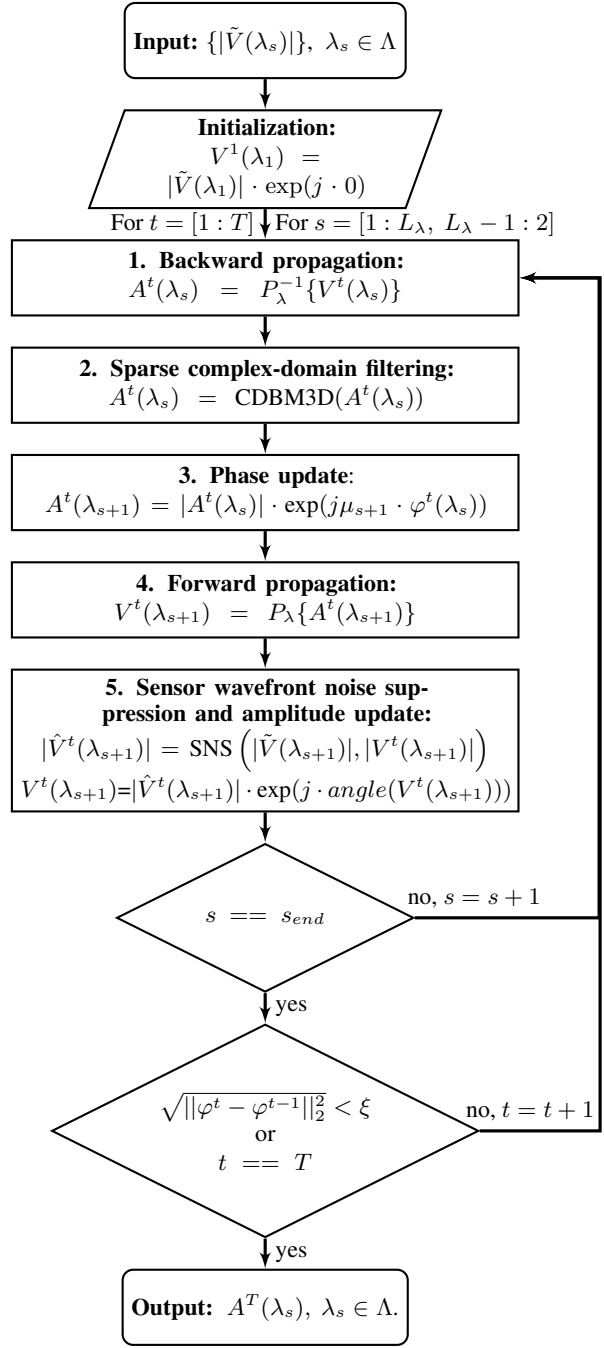
\begin{figure}[t!]
\centering
\begin{tikzpicture}[node distance=0.4cm]
\tikzstyle{startstop} = [rectangle,thick, rounded corners, minimum width=3cm, minimum height=1cm,text centered, draw=black] 
\tikzstyle{io} = [trapezium,thick, trapezium left angle=70, trapezium right angle=110, minimum width=2cm, minimum height=1cm, text centered,text width=3.5cm, draw=black] 
\tikzstyle{process} = [rectangle,thick, minimum width=2.5cm, minimum height=1cm, text centered, text width=6.5cm, draw=black] 
\tikzstyle{decision} = [diamond,thick, minimum width=1.6cm, text centered,text width=2.9cm, draw=black,aspect=2] 
\tikzstyle{arrow} = [thick,->, >=latex',  line width=1.2pt, shorten >=0.02pt] 

\node (start) [startstop]
{\footnotesize $\textbf{Input: }\{  |\Tilde{V}(\lambda_s)| \}\text{, }~\lambda_s\in\Lambda$};

\node (in1) [io, below=of start]
{\footnotesize \textbf{Initialization:}\\ 
$V^{1}(\lambda_1)=|\Tilde{V}(\lambda_1)|\cdot\exp (j\cdot 0)$};
	
\node (pro1) [process, below=of in1] 
{\footnotesize \textbf{1. Backward propagation:} \\ 
$A^{t}(\lambda_s)=P_{\lambda}^{-1}\{V^{t}(\lambda_s)\}$};

\node (pro2) [process, below=of pro1] 
{\footnotesize \textbf{2. Sparse complex-domain filtering:} \\ 
$A^{t}(\lambda_s)=\text{CDBM3D}(A^{t}(\lambda_s))$};


\node (pro3) [process, below=of pro2] 
{\footnotesize \textbf{3. Phase update}$\text{:}$ \\
$A^{t}(\lambda_{s+1})=|A^{t}(\lambda_s)|\cdot \text{exp}(j\mu_{s+1}\cdot {\varphi}^{t}(\lambda_s))$};
	
\node (pro4) [process, below=of pro3] 
{\footnotesize \textbf{4. Forward propagation:} \\ 
$V^{t}(\lambda_{s+1})=P_{\lambda}\{A^{t}(\lambda_{s+1})\}$};


\node (pro5) [process, below=of pro4] 
{\footnotesize \textbf{{5. Sensor wavefront noise suppression and amplitude update:}} \\$| \hat{V}^t(\lambda_{s+1})|=\text{SNS}\left(|\Tilde{V}(\lambda_{s+1})|,|{V}^t(\lambda_{s+1})|\right)$
\\
$V^{t}(\lambda_{s+1})$=$|\hat{V}^t(\lambda_{s+1})|\cdot \text{exp}(j\cdot angle(V^{t}(\lambda_{s+1})))$ \\
};
\node (dec1) [decision, below=of pro5] {\footnotesize $s==s_{end}$};

\node (dec2) [decision, below=of dec1] {\footnotesize $\sqrt{||{\varphi}^{t}-\varphi^{t-1}||_{2}^{2}} < \xi$\\or\\ $t==T$ };

\node (out1) [startstop, below=of dec2] 
{\footnotesize $\textbf{Output: }$   $A^{T}(\lambda_s){,~}\lambda_s\in\Lambda $.};
 

\draw[arrow] (start) -- (in1);
\draw[arrow] (in1) -- node[anchor=east] {\footnotesize $\text{For}~t=[1:T]$}  node[anchor=west] {\footnotesize $\text{For} ~s=[1:L_\lambda,~ L_\lambda-1:2 ]$}(pro1); 
\draw[arrow] (pro1) -- (pro2);
\draw[arrow] (pro2) -- (pro3);
\draw[arrow] (pro3) -- (pro4);
\draw[arrow] (pro4) -- (pro5);
\draw[arrow] (pro5) -- (dec1);
\draw[arrow] (dec1.east) node[anchor=south]{\footnotesize ~~~~~~~~~~~~~~~~~~~~no, $s=s+1$} -++  (2.45,0) |-   (pro1);
\draw[arrow] (dec1)  -- node[anchor=west] {\footnotesize yes}(dec2);
\draw[arrow] (dec2.east) node[anchor=south]{\footnotesize ~~~~~~~~~~~~~~~~~no, $t=t+1$} -++  (1.6,0) |-   (pro1);
\draw[arrow] (dec2)  -- node[anchor=west] {\footnotesize yes}(out1);	
\end{tikzpicture}	
\caption{HSPR algorithm flowchart.}
\label{Fig-flowchart}
\end{figure}
We present {\color{black} the phase retrieval algorithm in the context of thickness (profile) measurement for a transparent object with HS illumination in the self-reference configuration of the optical system.
The coherent laser beam of the wavelength $\lambda$ propagates through the object. It results in the phase delay, $\varphi _{\lambda }(x,y)$, between the input and output beams. 
The corresponding output wavefronts   $A_{\lambda }(x,y)=|A_{\lambda
}(x,y)|\cdot \exp (j\varphi _{\lambda }(x,y))$ are different for different wavelengths, $\lambda \in \Lambda $, where $\Lambda $ is
a set of the HS wavelengths. These wavefronts at the sensor plane are calculated as
{\color{black} $V(\lambda)={P_{\lambda}}\{A(\lambda)\}$, 
where ${P_{\lambda}}$} stays for the forward propagation operators depending on wavelength. 

{\color{black} For the finite discrete set of the shifts $\Delta z$ we can calculate the spectral distribution $|V(\lambda)|^2$ only for the corresponding finite set of the wavelength, i.e. for ~$\Lambda=[\lambda_1,...,\lambda_{L}]$. Then, integration in \eqref{eq:interferog}-\eqref{eq:amplitude} is replaced by summation and the integral Fourier transform by the discrete Fourier transform. 
}
Therefore, according to \eqref{eq:interferog} the measurements by the sensor are defined as 
\begin{equation}
J(z)=\sum_{\Lambda}\left|V(\lambda)\left(1+\exp (j \dfrac{2 \pi}{ \lambda} z)\right)\right|^{2}.
\label{eq:interferogdiscrete}
\end{equation}
This equation gives the intensity calculated over the spectrum range and registered by each pixel of the sensor.

The developed  Hyperspectral Phase Retrieval (HSPR) algorithm is composed of two separate stages. 
The first stage is the spectrum analysis defining the intensity spectral distributions  $|V(\lambda)|^2$ from the registered set of $J(z)$. These calculations are based on Discrete and Fast Fourier Transform (FFT), for the details of this part of the algorithm we refer to  \cite{katkovnik2019hyper_cosine}.

The second stage is phase retrieval, i.e. a reconstruction of the object complex transfer functions $A_{\lambda
}(x,y)$ from the already given spectra $|V(\lambda)|^2, \lambda\in\Lambda$.
A flowchart of the proposed phase retrieval algorithm is presented in Fig.~\ref{Fig-flowchart}.
The algorithm is iterative using typical forward/backward propagation between the object and sensor planes. The following points define the originality of this algorithm:

(1) At step 2, the complex domain CDBM3D filter is used for denoising of object transfer functions for each wavelength and in each iteration;

(2) At step 5, special filtering is produced for amplitudes of the wavefronts at the sensor plane.}

In the case of HS illumination for a transparent object, it is common to rely on the assumption that object wavefronts produced by neighboring wavelengths are similar \cite{Zhuang2018}. It is quite a natural assumption since neighboring wavelengths are close to each other and phase properties of the object are smooth functions of wavelengths.
We utilize this assumption and \eqref{eq:thinkness} to provide a connection between neighboring wavelength phases by introducing the coefficient  
\begin{equation}
 \mu_{s}=\dfrac{\lambda_{s-1}(n_{\lambda_{s}}-1)}{\lambda_{s}(n_{\lambda_{s-1}}-1)},
 \label{eq:mumumu}
\end{equation} 
such that $\varphi _{\lambda_s }=\varphi _{\lambda_{s-1}}\cdot \mu_{s}$. Here $n_{\lambda_{s}}$ is the refractive index of the object.

In the flowchart of the algorithm in Fig.~\ref{Fig-flowchart} as input, we have spectral amplitudes $|\Tilde{V}(\lambda_s)|$ at the sensor plane for the whole range of the HS wavelengths reconstructed at the first stage of the algorithm. 
 For the initialization of iterations, we create the first complex-valued wavefront $V^{1}(\lambda_1)$ using for the amplitude the first component of the HS $|\Tilde{V}(\lambda_1)|$ with zeros guess for the initial phase. 
 After the initialization, we start the two consecutive ``for'' loops with running variables $t$ and $s$ for cube and wavelength iterations, respectively. 
 For smooth cube iterations $t$, the loop on $s$ should run consequently through all wavelengths with start and stop at the first wavelength. 
 
 Step 1 is the backward propagation from the sensor plane to the object plane, as a result, we obtain an estimate for the object wavefront $A^{t}(\lambda_s)$. $P_{\lambda}$ is a wavelength-dependent propagation operator which is defined by the angular spectrum propagation model \cite{GOODMAN} and with the superscript `$-1$' it indicates the backward propagation.
 At step 2, we perform complex-domain sparse noise suppression in $A^{t}(\lambda_s)$ by the CDBM3D filter \cite{Katkovnik-2017-CDBM3D}, which processes amplitude and phase jointly and additionally to sparsity utilizes the correlation between amplitude and phase.  

At step 3, we make phase recalculation {\color{black} from $s$ to $s+1$ wavelength according to \eqref{eq:mumumu}.
This spectral update is based on the following speculations.
The phase delay of the object is varying according to the object thickness.
A link between the phase delay and object thickness is defined by the equation:
\begin{equation}
  h_{o}(x,y)   =\dfrac{\varphi _{\lambda }(x,y)\cdot\lambda}{2\pi\cdot (n_{\lambda }-1) }.
  \label{eq:thinkness}
\end{equation}
Here, $h_{o}(x,y)$ is a thickness of the transparent object at the point $(x,y)$.
}

 At step 4, we perform forward propagation of the object wavefront to the sensor plane, where at step 5 keeping the phase of $V^t(\lambda_{s+1})$  unchanged we make amplitude update by filtering the amplitude $|\hat{V}^t(\lambda_{s+1})|$. This noise filtering is performed by {\color{black} the Sensor Noise Suppression (SNS) algorithm. This algorithm is derived based on the optimal solutions for maximum likelihood criteria for  Gaussian and Poissonian noise distributions.
 Both solutions are published in \cite{Katkovnik-ArXive}. In this paper, we use the solution for the Gaussian noise.  
 The obtained optimal estimate of the amplitude locates between observations $|\Tilde{V_{s}}(\lambda)|$ and the amplitude of the object wavefront propagated to the sensor plane $|{V_{s}}^t(\lambda)|$. 
 
 Next, in decision block, the stopping rule is applied, defined by the maximum number of iterations on $t$ and differences between the phase  estimates in successive iterations.}

The phase retrieval part of the developed HSPR algorithm shown in Fig.~\ref{Fig-flowchart} has a structure similar to the algorithms proposed earlier for multispectral phase retrieval (MPR) in the papers ~\cite{Pedrini-2008,Pedrini-2012}. Let us discuss briefly differences between HSPR and those MPR  algorithms. First of all, different optical setups considered in this and the mentioned papers. The observations in \cite{Pedrini-2008,Pedrini-2012} are obtained in experiments  separate for each laser bands, when in this paper all wavelengths go through the object simultaneously. 
Second, while the algorithms MPR and HSPR look similar due to the standard iterative structure with forward and backward wavefront propagation as it is originated from Gerchberg and Saxton \cite{Gerchberg1969}, 
the efficient noise suppression in steps (2) and (5) define the originality of the developed algorithm HSPR.
These filtering steps enhance the HSPR algorithm and help to obtain  cleaner reconstructions with faster convergence since noise in the iterative algorithm is predisposed for corruption and ruin the reconstruction. 

  The noise suppression becomes extremely significant in the case of HSPR   because in the proposed scheme all spectrum components are observed simultaneously and, therefore, in presence of noise in one region of the spectrum it will be spread throughout the whole spectrum (Fellgett's disadvantage \cite{fellgett_1950}).

\section{Results}
\subsection{Simulations}
\label{sec:simulations}
 For a demonstration of the algorithm performance, we provide modeling of the HS data obtained by the system described in Section~\ref{sec:problem}. 
 We use the broadband light spectrum corresponding to a broadband supercontinuum laser source (see spectrum in Fig.~\ref{fig:scheme_and_spectra}(b)) and modeled hyperspectral data registration as interference observations according to \eqref{eq:interferogdiscrete} providing $N=2000$ shifts in the delay line which correspond to total $Z=200$~$\upmu$m with $\Delta z=100$~nm.  Since the laser spectrum is not uniform, in regions of the low laser intensity, signal-to-noise ratio (SNR) is low and these regions cannot be used for phase reconstruction. For our tests, we exploit the wavelengths from the high laser intensity region $[680:820]$~nm.   

\begin{figure}[t!]
\centering
\includegraphics[width=.7\linewidth]{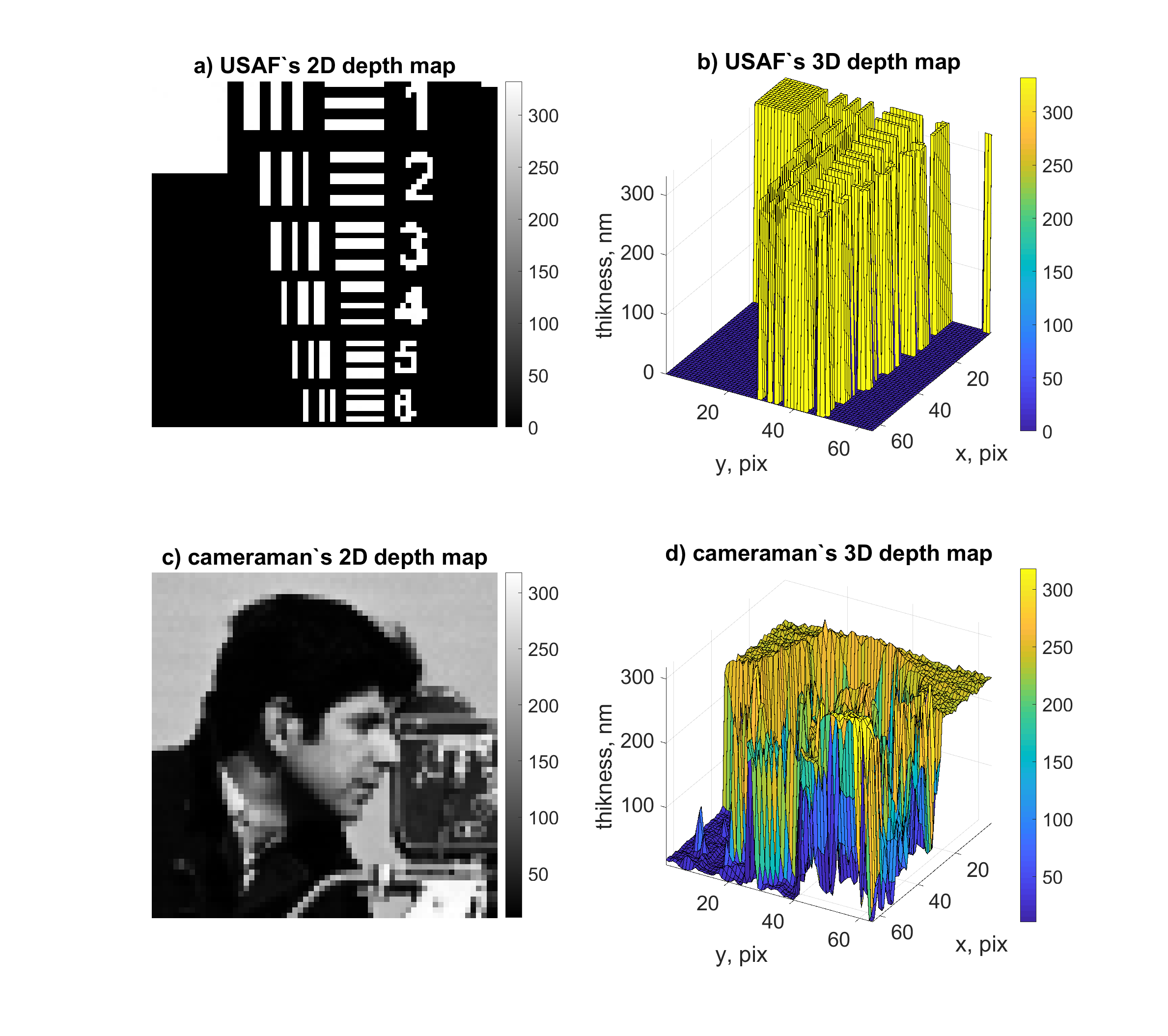}
\caption{Simulated objects surface maps. a) and b) are 2D and 3D USAF depth maps; c) and d) are 2D and 3D - Cameraman.   }
\label{fig:Objects}
\end{figure}

 As objects under investigation, we model the transparent phase objects assuming that objects'  transfer function $A_{\lambda }(x,y)=|A_{\lambda}(x,y)|\cdot \exp (j\varphi _{\lambda }(x,y))$ with the amplitude $|A_{\lambda}(x,y)|=1$ and the phase images $\varphi_{\lambda}(x,y)$ given as the USAF test-target and Cameraman test-image. The distributions of the object phase delays  for each $\lambda$ are calculated according to \eqref{eq:thinkness} with maximum $h_o=317$~nm.
 Figure \ref{fig:Objects} demonstrates the USAF (a-b) and Cameraman (c-d) phase objects  (depth distributions) as 2D and 3D images. The USAF test-target provides a binary object model, while the Cameraman test-image has a more complex structure with a multilevel piece-wise continuous  depth map. 
 
\begin{figure}[t!]
\centering
\includegraphics[width=0.7\linewidth]{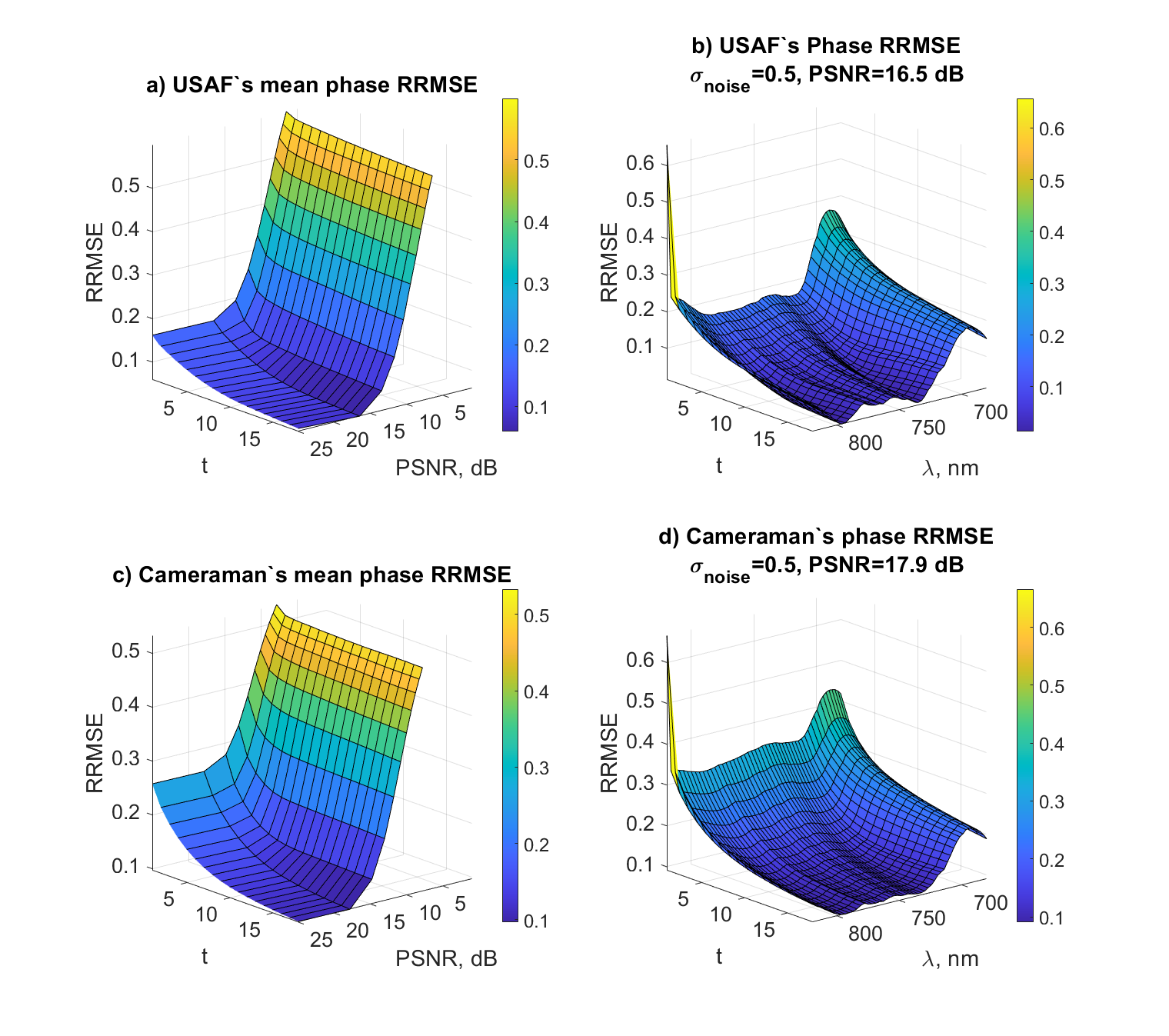}
\caption{RRMSE maps. 
a) USAF's mean RRMSEs depending on the number of iteration $t$ and noise level in the observations PSNR. b) RRMSEs for all wavelengths $\lambda$ of the USAF HS cube depending on iteration number $t$ for the case of PSNR$=16.5$~dB. c) Cameraman's mean RRMSEs depending on the number of iteration $t$ and noise level in the observations PSNR. d) RRMSEs for all wavelengths $\lambda$ of the Cameraman HS cube depending on iteration number $t$ for the case of PSNR$=17.9$~dB.  }
\label{fig:RRMSEs_maps}
\end{figure}
To check algorithm robustness, we include noise in our modeling. The noise $\varepsilon$ is additive zero-mean i.i.d. Gaussian  with the standard deviation $\sigma_{noise}$:
\begin{equation}
    J_{noisy}(z)=J(z)+\varepsilon(\sigma_{noise}). \label{eq:noise}
\end{equation}
The accuracy of phase reconstruction is defined by the Relative Root-Mean-Square-Error (RRMSE) criterion:

\begin{equation}
\text{RRMSE}_{\varphi }=\frac{\sqrt{||\hat{\varphi}_{est}-\varphi _{true}||_{2}^{2}}}{\sqrt{||\varphi _{true}||_{2}^{2}}}\text{, } \\
\label{eq:RRMSE}
\end{equation}%
where $\hat{\varphi}_{est}$ is the reconstructed phase and  $\varphi_{true}$ is the noiseless true phase, $||.||$ stays for the Frobenius norm.  
RRMSE values less than 0.1 correspond to a good quality phase reconstruction.

{\color{black} For characterization of the noise level at the sensor plane we introduce the Peak Signal-to-Noise Ratio (PSNR)  defined as 
\begin{equation}
    \text{PSNR}=10\log _{10}\left(\frac{\text{max$_{x,y,z}$}(J (z))}{\sigma_{noise}}\right),
    \label{eq:PSNR}
\end{equation}

where max$_{x,y,z}(J(z))$ is the maximum of the registered intensity of the   beam at the sensor plane calculated by maximization on $z$ and $(x,y)$, where $(x,y)$ are the spatial coordinate of $2D $ images; $\sigma_{noise}$ is the standard deviation of the additive noise.}

Figure~\ref{fig:RRMSEs_maps} demonstrates phase RRMSE for the object reconstructions.
In plots a) and c): 
RRMSE values depending on the noise level given by PSNR and the number of iteration $t$; in plots b) and d)  RRMSE values depending on wavelength $\lambda$  and  iteration number for the given noise $\sigma_{noise}=0.5$.

{\color{black} Descending RRMSEs for growing $t$ indicate consecutive improvement of the reconstruction from iteration-to-iteration.   
It can be concluded from RRMSE surfaces a) and c) that  reliable objects' reconstruction might be obtained only for low noise level with high      PSNR values ($>18$~dB) and iteration number $t>15$. It is caused by Felgett's disadvantage with noise leakage from one spectral component to others.} 
{\color{black} Surfaces b) and d) are provided for  $\sigma_{noise}=0.5$, the RRMSE values in the region of $\lambda=700$~nm reflect that the lower illumination intensity results in worse reconstruction quality.      
In general, the appearance of the RRMSE surfaces is similar for both objects under investigation, however, RRMSE values for USAF are lower than those for Cameraman. It appears that Cameraman is more complex for the considered algorithm than binary USAF.
}

\begin{figure}[t!]
\centering
\includegraphics[width=0.7\linewidth]{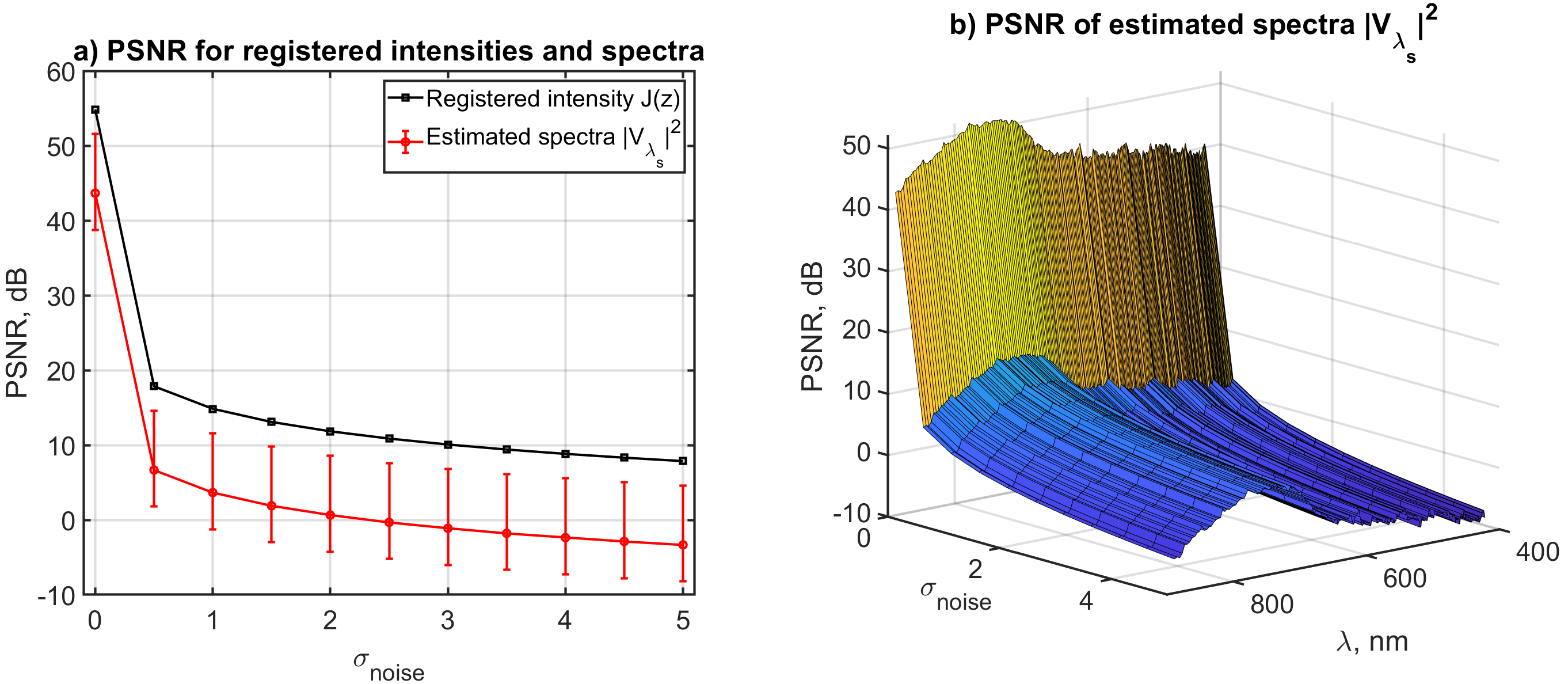}
\caption{a) PSNR dependencies from noise standard deviation $\sigma_{noise}$ for registered intensities $J(z)$, black squares curve, and for estimated spectra $|V_{\lambda_S}|^2$, red circles. Latter is presented as a mean value of PSNRs for the whole spectra $\Lambda$. Error bars correspond to the PSNR range for $\Lambda$. b) PSNR map of the estimated spectra  for different $\sigma_{noise}$ and wavelengths $\lambda$. }
\label{fig:noises_in_obser_and_spectra}
\end{figure}

Let us give some comments on the noise level in the intensity observations \eqref{fig:noises_in_obser_and_spectra} and in the spectra $|V(\lambda_s)|^2$. 
These spectra are calculated using the Fast Fourier Transform (FFT) applied to the sequence $J(z)$ as $|V(\lambda_s)|^2=FFT(J(z))/N$, where $N$ is a length of this sequence, i.e. a number of shifts in $z$. 
It follows, that the noise in these estimate of $|V(\lambda_s)|^2$ obtained by application of FFT to $J_{noisy}(z)$ is zero-mean white Gaussian, as in the observations but  with the standard deviation equal to $\sigma_{noise}/{\sqrt{N}}$, which is the same for all spectral estimates $|\Tilde{V}(\lambda_s)|^2$.
 
PSNR for observations is defined by  \eqref{eq:PSNR}, where in the nominator we have the maximum intensity calculated as the weighted sum of intensities of the spectral components. It follows that maximum spectral intensities $|{V}(\lambda_s)|^2$ different for different $\lambda$ take values lower than the observed intensity. For instance, if $|V(\lambda_s)|^2$ have close values for all $\lambda_s$ then we can say that approximately $|V(\lambda_s)|^2\sim J(z)/N$. 

Inserting this value in  \eqref{eq:PSNR}, as well as the obtained above value for the noise standard deviation for the spectral estimate, results in the following formula for PSNR calculated in the spectral domain
\begin{equation}
    \text{PSNR}_{|V_{\lambda_{s}}|^2}\sim10\log _{10}\left(\frac{\text{max$_{x,y,z}$}(J(z))}{\sigma_{noise}\cdot \sqrt{N}}\right).   \label{eq:PSNR_spectra}
\end{equation}

Then, PSNRs calculated for the spectral estimates  are smaller or much smaller than PSNR calculated for the observations. Thus, the noise level in the spectral domain is higher than that for the observations. It makes the problem of the noise removal for large $N$ quite demanding and of the crucial importance for qualitative phase imaging. 

This noise amplification phenomenon  in the spectral domain is illustrated in Fig.~\ref{fig:noises_in_obser_and_spectra} for our simulation data obtained for the Cameraman phase object.

As the spectra are varying as a function of wavelength, contrary to \eqref{eq:PSNR_spectra} we instead calculate PSNR according to the next formula:

\begin{equation}
    \text{PSNR}_{|V_{\lambda_{s}}|^2}=10\log_{10}\left(\frac{\text{max$_{x,y}$}(|V_{\lambda_{s}}|^2)}{\sigma_{noise}/{\sqrt{N}}}\right).   \label{eq:PSNR_spectra_varying_var}
\end{equation}
Here, the expression  in the brackets is a ratio of the spectrum intensity for the wavelength $\lambda_{s}$ to the standard deviation corresponding to the noise in the spectrum estimates. 
Thus, PSNR is evaluated for each spectral component independently. Note that maximization in this definition of  $\text{PSNR}_{|V_{\lambda_{s}}|^2}$ is restricted to the coordinates $(x,y)$.

In Fig.~\ref{fig:noises_in_obser_and_spectra}, we show the mean value of $\text{PSNR}_{|V_{\lambda_{s}}|^2}$ calculated for the whole range of the spectra as well as the error bar showing variations of $\text{PSNR}_{|V_{\lambda_{s}}|^2}$ from minimal to maximal values. It is seen that PSNR in the spectral domain is much lower than that for the observations and this statement holds for all values of $\sigma_{noise}$.
It once more illustrates the noise amplification effect appeared in HS phase imaging.
This effect is the interpretation of the well known in Fourier spectroscopy Felggett's disadvantage \cite{fellgett_1950}.

\begin{figure}[b!]
\centering
\includegraphics[width=0.7\linewidth]{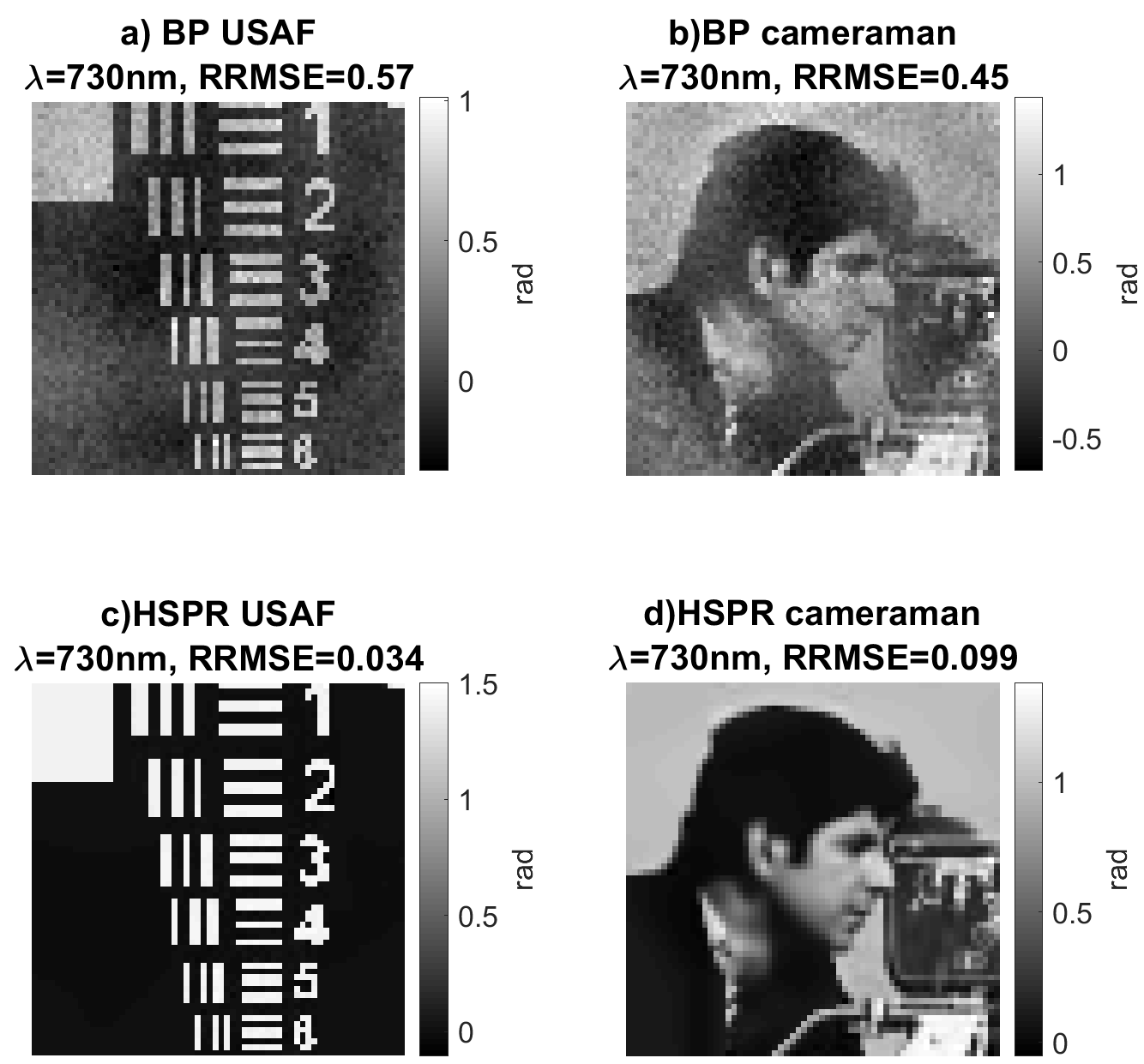}
\caption{Phase reconstruction results for back propagation (BP) (a and b) and HSPR (c and d), USAF and Cameraman objects, respectively, $\lambda=730$~nm.
}
\label{fig:reconstruc_simulations}
\end{figure}

Figure~\ref{fig:reconstruc_simulations} illustrates the crucial importance of iterations and filtering embedded in the developed algorithm.
In Figs.~\ref{fig:reconstruc_simulations}(a-b) we show the results obtained by backward propagation of the spectral intensities obtained by FT from noisy observations. Actually, it is the initial  iteration of our algorithm giving the first approximation of the objects.
The comparison with the corresponding HSPR reconstructions in Figs.~\ref{fig:reconstruc_simulations}(c-d) is a clear demonstration of the performance of the algorithm and efficiency of its iterations.

\subsection{Physical experiments}
\label{sec:experiment}
 
\begin{figure}[b!]
\begin{minipage}[h]{0.38\linewidth}
\center{\includegraphics[width=1\linewidth]{{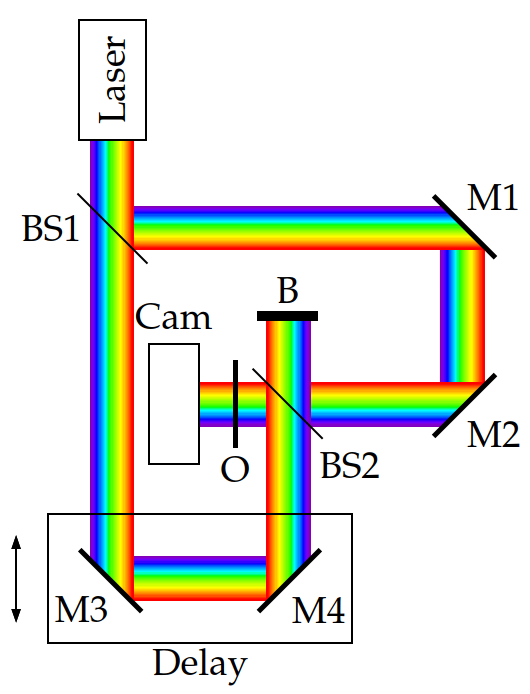}} \\ a)}  
\end{minipage}
 \hfill
\begin{minipage}[h]{0.5\linewidth}
\center{\includegraphics[width=1\linewidth]{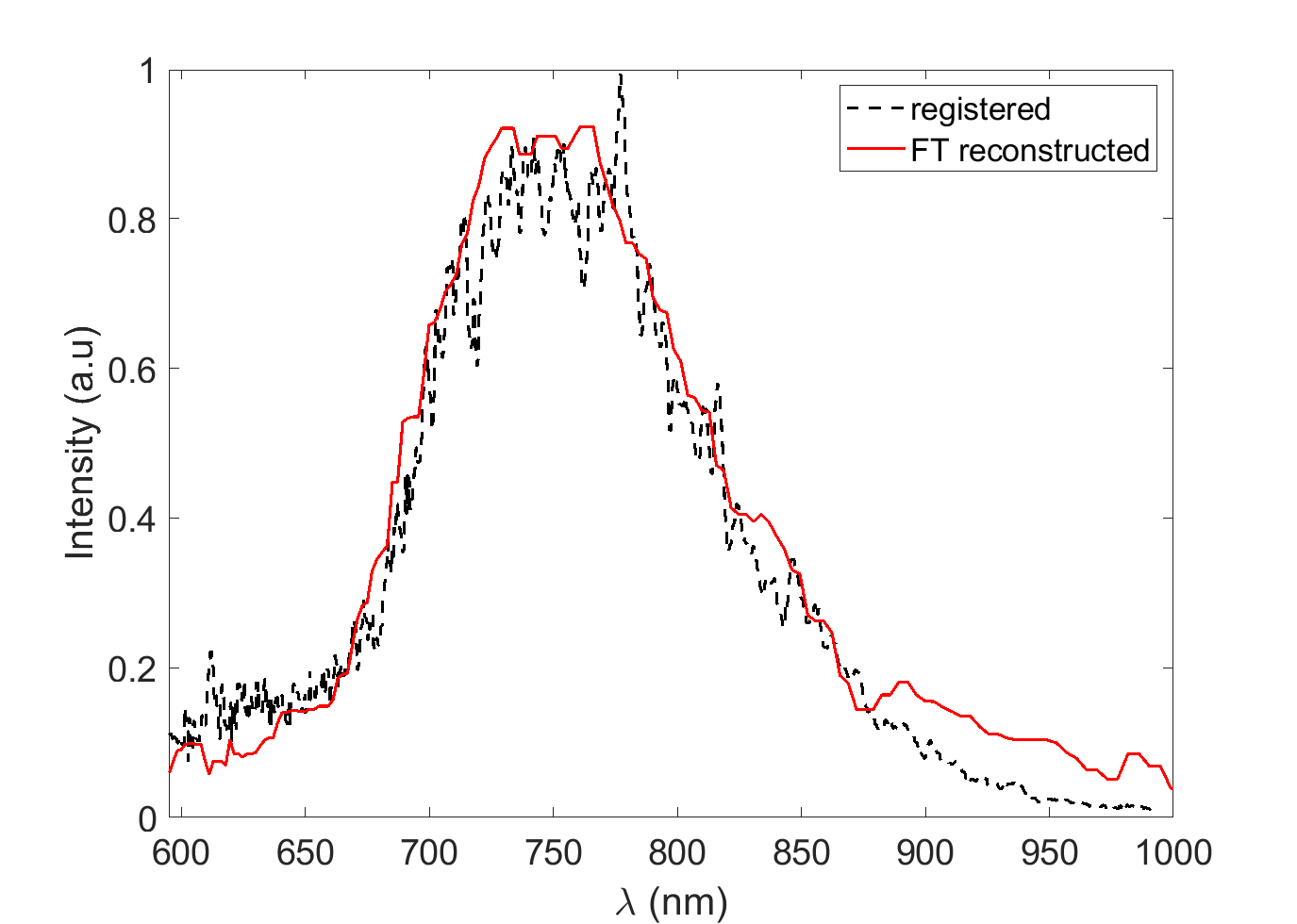} \\ b)}
\end{minipage}
\caption{(a) Hyperspectral phase retrieval setup. BS1-2 are beamsplitters, M1-4 are mirrors, ``O'' is a transparent object, ``Cam'' is a registering sensor, ``B'' is a light blocker. ``Delay'' is a moving delay stage. (b) Used spectrum: a black dash curve is a registered spectrum by a spectrometer and multiplied by camera quantum efficiency, a red solid curve is for the Fourier transform reconstructed spectrum.}
\label{fig:scheme_and_spectra}
\end{figure}

We have developed the HS lensless self-reference optical system shown in Fig.~\ref{fig:scheme_and_spectra}(a). 
The super-continuum laser, $\Lambda=470-2400$~nm (YSL photonics CS-5), is used as a source of light. The light beam is split equally into two beams by beamsplitter BS1. The first beam reflected by the mirrors M1 and M2 goes undisturbed to the beamsplitter BS2. The second beam goes through the delay line M3-M4, obtaining the different optical paths with respect to the first beam. 
BS2 merges the beams together and after BS2 beams go through the object ``O'' to the registering camera ``Cam'' (FLIR Chameleon, $2448\times2048$, pixel size $3.45~\upmu$m).  

The delay line is a piezo-based stage (Thorlabs NFL5DP20).
The step-size of this stage defines the spectral resolution of the setup: the minimal step of the delay line, $\Delta z$, should be at least twice smaller than the smallest wavelength, and the total moving distance of the delay line, $Z$, defines the spectral resolution for wavenumber $k$ as $\Delta k=\frac{1}{2\cdot Z}=\frac{1}{2 \cdot \Delta z \cdot N}$, here $Z=N\cdot\Delta z$ and $N$ is a number of steps of the stage.

\begin{figure}[t!]
\centering
\includegraphics[width=0.7\linewidth]{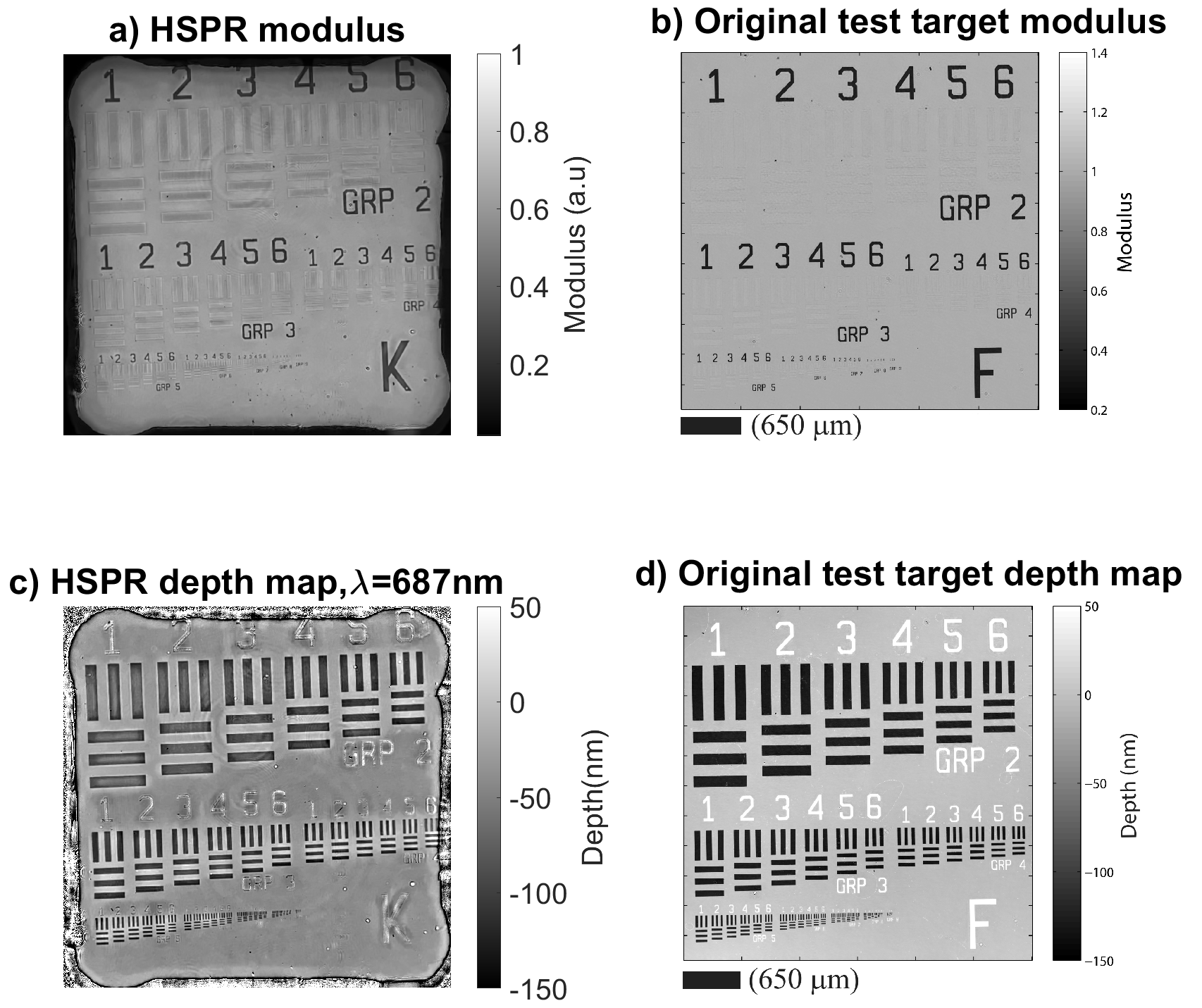}
\caption{ Test target images. a) Modulus and c) depth map reconstructed by HSPR for slice $\lambda=687~$nm; b) modulus and d) depth map provided by the manufacturer.}
\label{fig:experim1}
\end{figure}

In the provided experiments, the system parameters were $\Delta z = 59.7$~nm  and $N=1880$, that correspond to $\Delta k=44.6$~cm$^{-1}$.
The distance between the object and the camera was $d=16$~mm. 
The used laser spectrum is wide, however, due to camera sensitivity we utilize only part of the spectral illumination, as it is shown in Fig.~\ref{fig:scheme_and_spectra}(b), the black dash curve demonstrates laser spectrum registered by a spectrometer (Thorlabs CCS200) and multiplied by quantum efficiency of the registration camera. The red curve in Fig.~\ref{fig:scheme_and_spectra}(b) is the spectrum $|V(\lambda)|$ reconstructed by Fourier transform \eqref{eq:amplitude}. This reconstruction is in a good agreement with the true spectral curve registered by the spectrometer.
As a test object, we used PhaseFocus test target \cite{PhaseFocus_2016}, which modulus and depth map are presented in Fig.~\ref{fig:experim1}(b,d), respectively, it is a phase object with an etched surface in fused silica glass, the depth of the etched features is 127~nm. The smallest groups from 6 to 9 have the smallest feature sizes of $10,~6.5,~3.5,~2.0~\upmu$m, respectively.

\begin{figure*}[t!]
\centering
\includegraphics[width=1\linewidth]{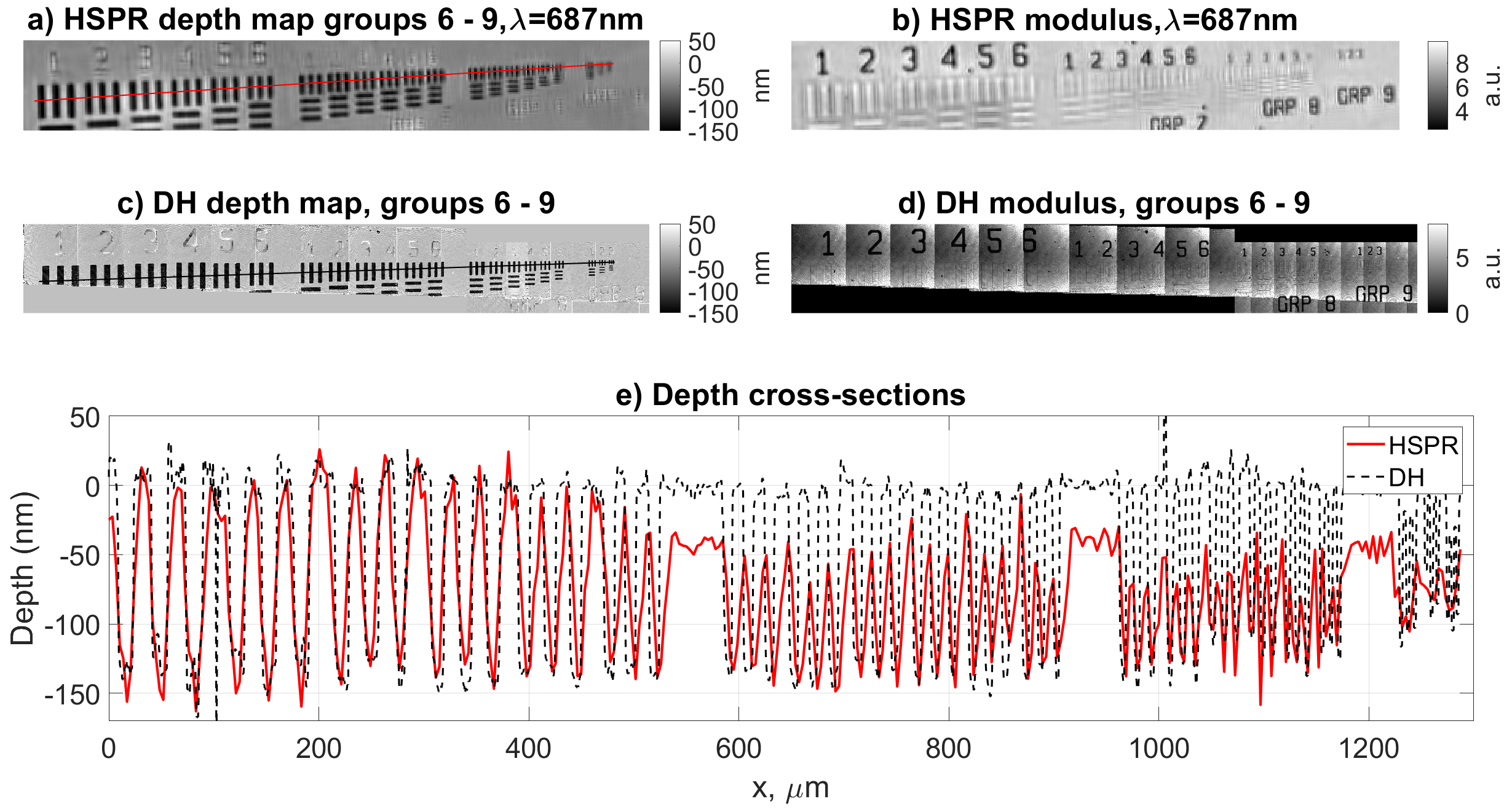}
\caption{Reconstruction results for the smallest 6 - 9 groups of the object. a) Depth map reconstructed by HSPR for slice $\lambda=687~$nm; b) HSPR modulus for slice $\lambda=687~$nm; c) Depth map obtained by digital holography (DH); d) DH Modulus; e) Longitudinal cross-sections corresponding to lines from images (a) and (c), red solid curve is for HSPR $687$~nm, black dash curve is for DH.}
\label{fig:experim_merging}
\end{figure*}

Reconstruction results by the HSPR algorithm are presented in Fig.~\ref{fig:experim1}(a,c), where modulus and depth map of the phase test target corresponding to $\lambda=687$~nm are demonstrated. 
From these images, it is seen that HSPR provides good quality reconstructions for both modulus and phase for the whole test target.
The HSPR reconstructed modulus image in Fig.~\ref{fig:experim1}(a) is coincide with the image provided by the manufacturer (Fig.~\ref{fig:experim1}(b)) with clearly resolved modulus features and almost not distinguishable features for phase regions. 
 The difference of letters ``K'' and ``F'' in the bottom right corner is explained by the manufacturer as they provide the same image with letter ``F'' for all customers, however, each test target has its own distinguishing letter (``K'' in our case). Other parameters are the same.  

For clarity of the presented results, we provide Fig.~\ref{fig:experim_merging} where the smallest groups 6,7,8,9 of the object presented and compared with depth and modulus reconstructed by digital holography (DH).
Figure~\ref{fig:experim_merging}(a,b) shows the depth map and modulus of the object corresponding to $\lambda=687$~nm; Fig.~\ref{fig:experim_merging}(c, d) - DH reconstruction, and Fig.~\ref{fig:experim_merging}(e) - a plot of depth cross-sections for the groups 6-9 reconstructed by HSPR (red solid curve) and DH (black dash), cross-sections are taken from lines with the same colors from Fig.~\ref{fig:experim_merging}(a) and Fig.~\ref{fig:experim_merging}(c), respectively. 
It is seen from the images that HSPR reconstructions correspond to the DH reconstructions and provided cross-sections show that reconstructed depth values equal to the real ones up to the elements 6 of the group 6 of the test target (region of 450 $\upmu$m of cross-sections in  Fig.~\ref{fig:experim_merging}(e)), which corresponds to $10~\upmu$m. 
Additionally, it is seen that the first four elements from group 8 appeared spatially resolved in terms of the Rayleigh criterion, but phase values are not correct. In that case, we may conclude that the given setup provides quantitative and qualitative phase estimations down to $10~\upmu$m resolution, and only qualitative results down to $4.5~\upmu$m.

Note, that the used for comparison DH system provides reference results based on a different optical setup with the use of lenses and holographic phase reconstruction algorithm \cite{Shevkunov_ICI_2016}. As compared with the system studied in this paper, disadvantages of the reference DH system are in utilization of objective with the magnification of $40\times$ and therefore an extremely small field of view $\approx0.15~$mm$^2$ (even for provided small image of groups 6-9 (Fig.~\ref{fig:experim1}(c)) we made stitching of 28 frames) and in lack of spectral information. 

\section{Conclusion}
A newly developed approach for hyperspectral phase retrieval in lensless self-referenced optical setup is introduced.
It is based on the principles of the Fourier Transform spectroscopy and iterative phase retrieval algorithms. 
The essential part of the algorithm is developed for  noise suppression based on sparse wavefront approximations. 
The provided approach is general for phase retrieval with HS illumination and can be applied for various self-reference schemes as for example \cite{Candeo_HSmicroscope_2019}, where self-referencing obtained by a birefringent delay line \cite{Brida:12}. We provided a transmissive setup for investigation of transparent objects, however, this approach is applicable and  for HS relief investigations in reflective setups. 
\section{Funding Information}

Jane and Aatos Erkko Foundation and Finland Centennial Foundation: Computational Imaging without Lens (CIWIL) project.

\section{Disclosures}

The authors declare no conflicts of interest.

 \bibliographystyle{unsrt} 

\end{document}